\documentclass[prl,aps,twocolumn,superscriptaddress]{revtex4-1}
\usepackage{latexsym,amssymb,amsfonts,amsmath,graphicx,bbm,color,bm,times,hyperref,amstext,relsize}

\begin{document}

\title{A quantum optical valve in a nonlinear-linear resonator junction}

\author{Eduardo Mascarenhas}
\affiliation{Departamento de F\'isica, Universidade Federal de Minas Gerais, Belo Horizonte, MG (Brazil)}

\author{Dario Gerace}
\affiliation{Dipartimento di Fisica, Universit\`a di Pavia, via Bassi 6, I-27100 Pavia (Italy)}

\author{Daniel Valente}
\affiliation{Institut N\'eel-CNRS, 25 Rue des Martyrs, 38042 Grenoble Cedex 9, Grenoble (France)} 
\affiliation{Universidade Federal de Mato Grosso, Cuiab\'a, MT (Brazil)}

\author{Simone Montangero}
\affiliation{Institut fur Quanteninformationsverarbeitung, A.-Einstein-Allee 11, D-89069 Ulm (Germany)}

\author{Alexia Auff\`eves} 
\affiliation{Institut N\'eel-CNRS, 25 Rue des Martyrs, 38042 Grenoble Cedex 9, Grenoble (France)}

\author{M. Fran\c{c}a Santos}
\affiliation{Departamento de F\'isica, Universidade Federal de Minas Gerais, Belo Horizonte, MG (Brazil)}

\date{\today}

\begin{abstract}
Electronic diodes, which enable the rectification of an electrical energy flux, have 
played a crucial role in the development of current microelectronics after the invention 
of semiconductor p-n junctions. 
Analogously, signal rectification at specific target wavelengths
has recently become a key goal in optical communication and signal processing.
Here we propose a genuinely quantum device with the essential rectifying features being 
demonstrated in a general model of a nonlinear-linear 
junction of coupled resonators. It is shown that such a surprisingly simple structure is a
versatile valve and may be alternatively tuned to behave as: 
a photonic diode, a single or two-photon rectified source turning a classical input into a 
quantum output depending on the input frequency, or a quantum photonic splitter. 
Given the relevance of non-reciprocal operations in integrated circuits, the nonlinear-linear
junction realises a crucial building component in prospective quantum photonic applications.
\end{abstract}


\maketitle

The electrical diode in a semiconductor p-n junction is the prototype of a rectifying 
device that  allows non-reciprocal electronic transport, which is key to information processing in integrated 
circuits~\cite{haus_book}; a number of applications currently require the realization of devices enabling 
unidirectional energy transport, from thermal~\cite{firstTD,chang2006sci,scheibner} and 
acoustic~\cite{BLiang2009prl,BLiang2010nmat} rectifiers, to all-optical diodes~\cite{gallo2001,yu2009nphot}.
The latter have now been attained in different configurations on-chip~\cite{Bi2011nphot,Lifan2012,Lira2012},
although always at the level of classical (i.e. many photons) signal transmission. 
A quantum optical rectifier may be generally defined as a two terminal, spatially nonreciprocal 
device that allows unidirectional propagation of single- or few energy quanta at a 
fixed signal frequency and amplitude. 
This conceptual extension of the classical diode operation can be foreseen 
to be important in the context of future applications in, e.g., integrated quantum 
photonics \cite{qp_review}, where novel quantum devices 
as single-photon transistors \cite{chang07np} and interferometers \cite{gerace_josephson}
have already been proposed, and where tunable rectification of quantum states is likely to 
play a role analogous to electrical diodes in current microchips.
However, proposals for quantum optical rectification have been quite limited in the literature,
to date. A device relying on non reciprocity induced by an external magnetic field in a 
doubly polarized waveguide has been proposed as a single-photon diode~\cite{shen2011prl},
where only conditional non-reciprocity depending on the specific polarization of the input 
state was shown.
Unconditional quantum optical diodes and transistors have
also been introduced in the context of atomtronic circuits \cite{AtDiode}, where an analogy
between one-dimensional optical lattices with cold atoms and electronic circuits was exploited
to suggest equivalent atom-based circuits over many sites.
 
In this letter we go beyond previous works, and describe a general scheme for a 
quantum optical device that works as an unconditional rectifier, elaborating on the simple analogy between 
the traditional p-n junction in semiconductor physics and a single nonlinear-linear (n-l) junction 
of two coupled resonators, as the building block of an elementary quantum optical valve. 
In particular, we show that this valve can be tuned to control energy transport at the quantum level
with direct applicability to current quantum technologies. Under a continuous monotonic pump the junction 
behaves as a rectified single or two photon source, depending on the input frequency, thus turning 
a classical input into a quantum output. 
At difference with previous investigations, we will use second-order photon correlation as a probe of
the quantum behavior of the device: a single-photon rectified source will be characterized by 
sub-Poissonian counting statistics in the transmitted signal, while a two-photon rectified source by
its super-Poissonian one. 
We also show that the junction behaves as a diode for fully quantum two-photon Fock states,
such that one photon activates the nonlinearity while the second photon is rectified. 
Finally, at high coupling between the resonators the junction splits the initial Fock state 
sending the two photons in opposite directions, thus acting as a quantum state splitter.

\begin{figure}
\includegraphics[width=\linewidth]{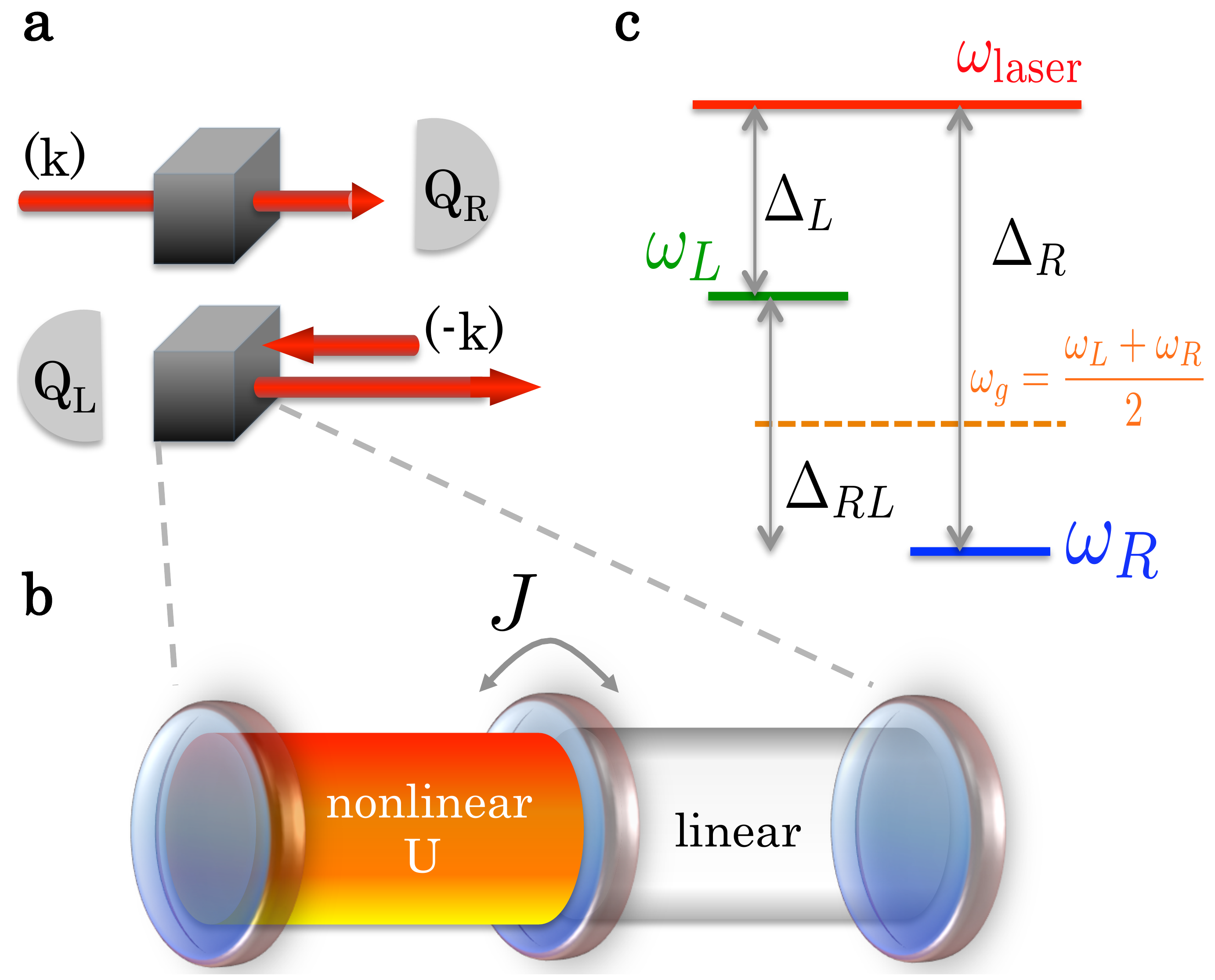}
\caption{ \textbf{Nonlinear-linear resonators junction} (a) Pictorial representation of a right-rectifying ``black box'' being pumped from left to right and then from right to left, with the transmission being significantly higher to the right. (b) Representation of light confining coupled resonators, one of which embedded in a nonlinear medium. (c) The representation of the frequencies and detunings of the driving laser and the resonators.}\label{Figure1}
\end{figure}

\textit{The model---}
We start from the concept of generic wave diodes in a nonlinear chain of  resonators~\cite{casatiGPD}:
a transmitted intensity at fixed incident amplitude and at the same frequency should be sensibly different in 
the two opposite propagation directions. 
To transfer these concepts to the quantum regime, we assume a chain of tunnel-coupled nonlinear sites, 
which can be generally described by the Bose-Hubbard model with single-particle interactions of the Kerr-type.
We will specify our treatment to a minimal two-site Bose-Hubbard Hamiltonian ($\hbar=1$)
\begin{equation}\label{eq:BHH}
H_{\textrm{n-l}}=\Delta_La^{\dagger}_La_L+\Delta_Ra^{\dagger}_Ra_R+\frac{U}{2}a^{\dagger}_La^{\dagger}_La_La_L+ J (a^{\dagger}_La_R+a_La_R^{\dagger}) \, ,
\end{equation}
such that the operator $a_i$ is the annihilation operator for the quanta in the $i$-th site, $U$ is the left site inter-particle interaction, 
and $J$ is the coupling strength between the two sites (usually determined by evanescent tunnel-coupling), 
which describes a two-site (left $L$, and right $R$) n-l junction as schematically shown in Fig. (\ref{Figure1}).
We notice that the generality of such a model has been established by effectively describing a wide variety of physical systems, 
from cold atoms in optical lattices~\cite{BHtheory,BHexperiment}, 
to strongly interacting photonic systems made with atoms coupled to optical resonators~\cite{Imamoglu,hartmann06} or optical fibres~\cite{chang08np}, 
spin chains~\cite{angelakis07}, arrays of superconducting circuits~\cite{leib2010njp}. 
or in open photonic devices, such as coupled arrays of nonlinear solid-state cavities~\cite{carusotto2009}. 
In the latter case, the on-site inter-particle interaction can be given by strong radiation-matter coupling of a 
single qubit-cavity system~\cite{nissen2012}, by enhanced Coulomb interaction of electron-hole pairs in semiconductor 
elementary excitations~\cite{ciuti06prb,carusot2010epl}, 
or by enhanced native nonlinearity of the bulk material thanks to the strong field confinement~\cite{ferretti2012}. 
In the case of weak nonlinearities, quantum interference between coupled modes can be exploited to reduce
the final modeling of the system to an effective Hubbard model in Eq. \ref{eq:BHH}, as proposed in Refs. \onlinecite{savona10prl,bamba}.
In the latter case, applications would imply fully passive quantum photonic devices compatible with standard materials
employed in optoelectronics \cite{ferretti2013}.

Owing to its out-of-equilibrium nature, the system dynamics is necessarily described by a balance between driven-dissipative terms, 
as it is typical of quantum optical systems~\cite{AtDiode,gerace_josephson,nissen2012}.
Either the left or right site of the junction can be coherently pumped, which is described by the Hamiltonian 
$H_p=F_ia^{\dagger}_i+F_i^{\ast}a_i$, with $F_i$ being the driving strength, where
$H_{\textrm{n-l}}+H_p$ is written in a reference frame rotating with the pumping frequency, $\omega_{\mathrm{laser}}$, 
with $\Delta_i=\omega_i-\omega_{\mathrm{laser}}$ such that $\omega_i$ is the $i$th site characteristic frequency. 
The level configuration is schematically shown in Fig. (\ref{Figure1}-c). 
 
We assume that the cavities (or sites) incoherently dissipate energy at a rate $\gamma$ determined by the openness of each site into the output channels (e.g., one-dimensional waveguides), and we define the output currents as number of quanta emitted per unit time from each site, i.e. 
$q_i(t)=\gamma\langle a^{\dagger}_i a_i\rangle(t)$. 
Formally, the average number of quanta emitted from the $i$th cavity during a time interval $\Delta t=t_2-t_1$, which is ideally the number of ``clicks'' registered as a photo-current in a single-photon detector, can be expressed as~\cite{QMaC,QNoise} 
%
\begin{equation}
{Q_i}(t_2,t_1)=\int_{t_1}^{t_2}q_i(t)dt   \, ,
\end{equation}
and the time-dependent quantum state of the two-site system is determined by the Liouville-von Neumann equation in Lindblad form~\cite{QMaC,QNoise,carmichael_book}
\begin{equation}\label{master:eq}
\dot{\rho}=\mathcal{L}(\rho)=-i[H_p+H_{\textrm{n-l}},\rho]+\mathcal{L}_L(\rho)+\mathcal{L}_R(\rho) \, ,
\end{equation}
with 
\begin{equation}\mathcal{L}_i(\rho)=-\frac{\gamma}{2}[a^{\dagger}_ia_i\rho+\rho a^{\dagger}_ia_i-2a_i\rho a^{\dagger}_i],\end{equation}
describing the energy dissipation from each site. 
Note that Eq.(\ref{master:eq}) faithfully describes any physical implementation of this model in the standard Markovian 
open system formalism~\cite{QMaC,QNoise}. 
If, for experimental or practical reasons, the output channels of the system are waveguides, the model properly describes 
the physical scenario in which these waveguides are independent and broadband.

With the dynamics and measurement processes specified, we define the rectifying factor
as the normalized difference between the output currents when the chain is pumped through the left and right resonator (indicated by the wave vectors $k$ and $-k$, respectively)
\begin{equation}
\mathcal{R}=\frac{{Q_{R}}[k]-{Q_{L}}[-k]}{{Q_R}[k]+{Q_L}[-k]},\label{RectFactor}
\end{equation}
such that the $\mathcal{R}$ factor measures the absolute rectification of the system: $\mathcal{R}=-1$ indicates maximal rectification with enhanced transport to the left (left rectification), $\mathcal{R}=0$ indicates no rectification, while $\mathcal{R}=+1$ indicates maximal rectification with transport to the right (right rectification). We also define the transport efficiency which is the amount of light that is transported to the desired direction. The transport efficiency to the right is given by 
\begin{equation}T_{R}=\frac{Q_R[k]}{Q_R[k]+Q_L[k]}.\end{equation}
Left efficiency $T_{L}$ is given analogously by interchanging $R$ with $L$ and $k$ with $-k$. We notice that the photo-detection time interval, $\Delta t$, can be taken as arbitrarily small in continuous-wave pumping and steady state regime, but it should be taken large enough in case of pulsed excitation in order to fully integrate the emitted pulses.

As an effective probe of quantum nonlinear features of this device, we calculate the photon counting statistics of the emitted light. This is defined by the second order correlation function at zero time-delay,  which is an experimentally relevant quantity and can be measured in a Hanbury Brown-Twiss (HBT) set-up with two single-photon detectors and a beam splitter~\cite{hbt_original}, theoretically given by
\begin{equation}
g^{(2)}_i(\pm k)=\frac{\langle a^{\dagger}_ia^{\dagger}_ia_ia_i\rangle_{(\pm k)}}{\langle a^{\dagger}_ia_i\rangle_{(\pm k)}^2} \, .
\end{equation}
This function gives values below unity for antibunched and above unity for bunched photons, respectively~\cite{carmichael_book}. 
Antibunching corresponds to a reduced probability that two photons are detected in coincidence at a given time, while it is the opposite for bunching.

\begin{figure}
\includegraphics[width=0.48\textwidth]{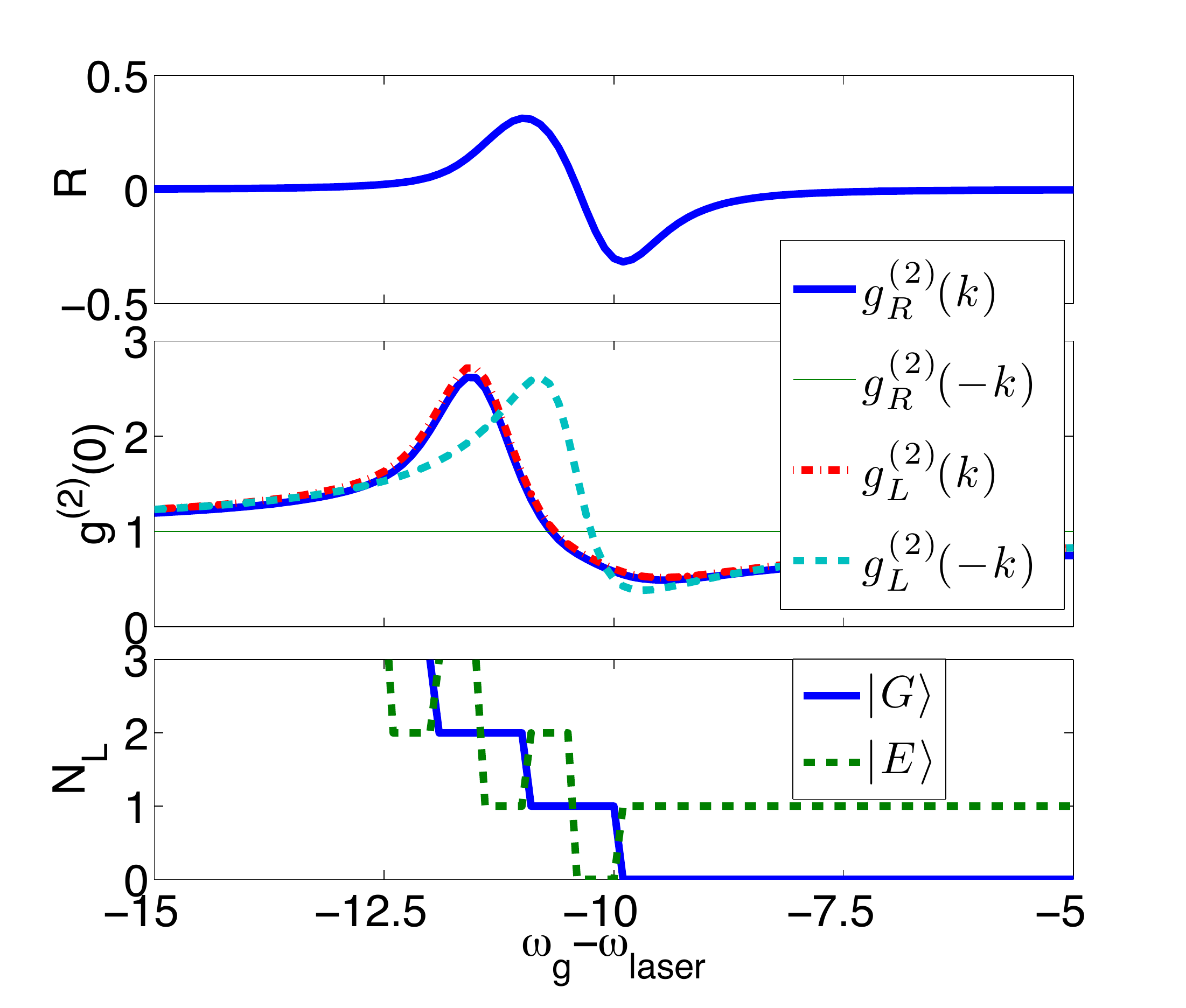}
\caption{\textbf{Non-equilibrium rectification under continuous-wave pumping, and the corresponding equilibrium excitations of the low-lying eigenstates probed by tuning the laser frequency.}
We show the rectification factor $\mathcal{R}$ (top panel), the second-order coherence function $g^{(2)}$ of the output light in both directions (middle), and the number of excitations in the left resonator, $N_L=\langle a^{\dagger}_La_L\rangle$, for the ground and first excited states of the bare hamiltonian (1) (bottom).
Parameters are: $U=\gamma$, $J=0.1\gamma$, and $F=0.5\gamma$. We assume $\Delta_{{RL}}=20\gamma$ with $\Delta_L=\Delta_{RL}/2+\omega_g$ and $\Delta_R=-\Delta_{RL}/2+\omega_g$.}\label{Figure2}
\end{figure}

\textit{Results---} Let us initially focus on Fig. (\ref{Figure2}), where we show the low $U/ \gamma$ regime of the system under continuous pump, for which we probe the n-l junction by scanning the laser frequency for a fixed $\Delta_{RL}$.
First, we turn our analysis to the full equilibrium quantum picture of the low lying eigenstates of the Hamiltonian \ref{eq:BHH}, more specifically focussing on the nonlinear cavity. This is justified for two reasons: at this point we will address the regime of low tunnelling between cavities and low pump intensities. Therefore, only the low photon states can be probed and only if they are close to resonance with the laser, and due to the low tunnelling the nonlinear effects can be directly associated to the states of the nonlinear resonator.
In Fig. (\ref{Figure2}-a) a maximum of left rectification corresponds to the condition $\omega_g-\omega_{\mathrm{laser}}=-10\gamma$, which simultaneously shows anti-bunched emission of the left site occurring exactly at the left cavity bare resonance (higher nonlinearities lead to stronger anti-bunching) in (\ref{Figure2}-b). At the same time, in the equilibrium picture the population of the left resonator in the global ground state, $N_L=\langle a^{\dagger}_La_L\rangle$,  switches from zero to one [see Fig. (\ref{Figure2}-c)] while the population of first excited state switches from one to zero ($|0\rangle \leftrightarrow|1\rangle$), thus showing that this process is predominantly a single photon process (as further confirmed by the anti-bunching statistics). In this case the junction turns the classical input into a quantum output, working as a rectified single photon source. On the other hand, the right rectifying process (bunched light) is predominantly a two photon process (with resonance condition $\Delta_L+U\approx0$, where the nonlinearity compensates for the detuning), as it can be observed in the low eigenstates excitations, where there is a switching from one-photon to two-photons state ($|1\rangle \leftrightarrow|2\rangle$). With this low lying states analysis we can fully connect the equilibrium properties of the system with its non-equilibrium response as a driven-dissipative quantum diode for one and two photons.

\begin{figure}
\includegraphics[width=\linewidth]{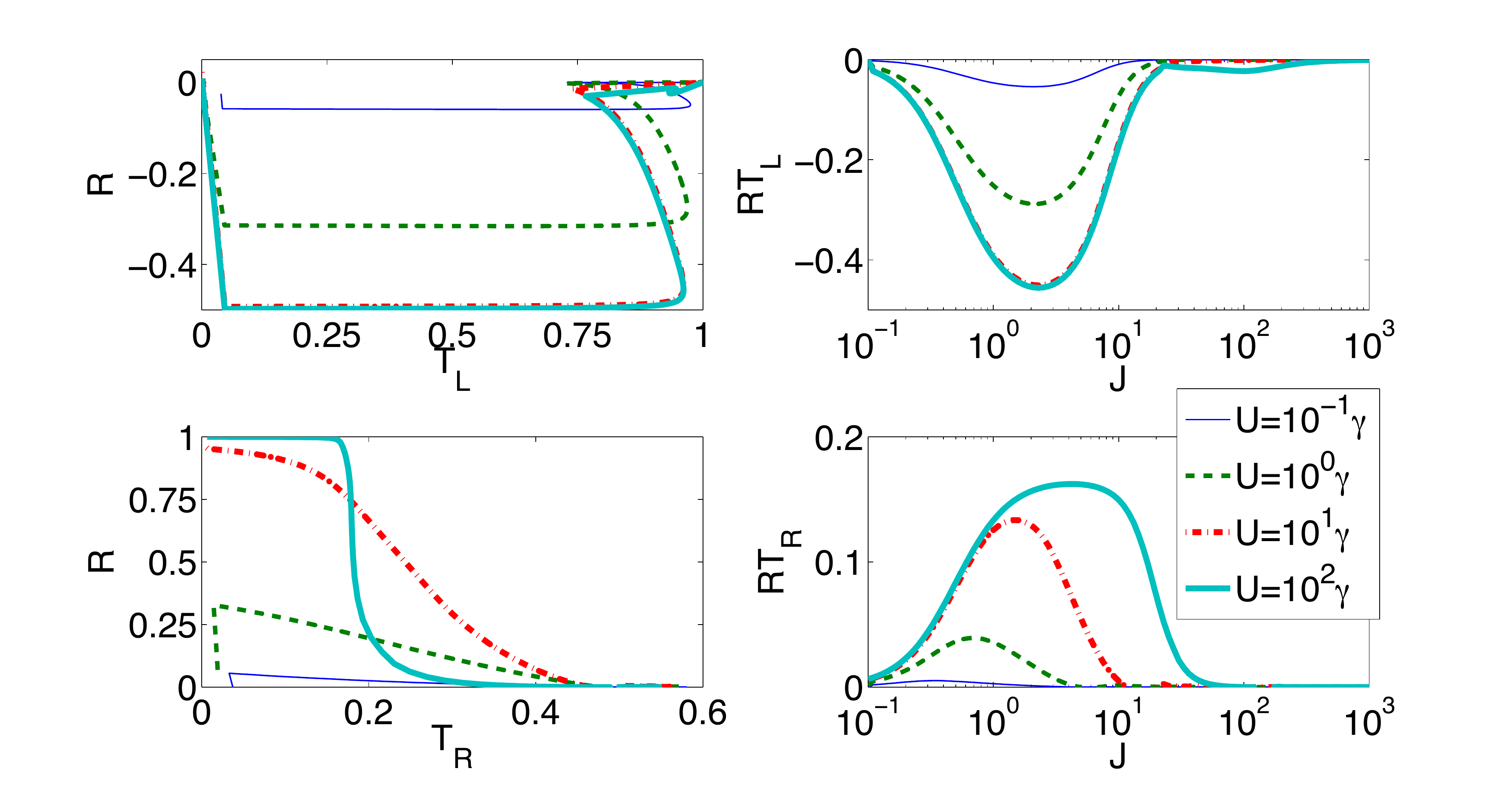}
\caption{\textbf{Rectification and the corresponding transport efficiencies with frequency optimization.}
(Top Left) Parametric plot of the optimized rectification and transport efficiency to the left and (Top right) their corresponding product as a function of the resonators coupling with $\Delta_{RL}=20\gamma$. (Bottom Left) Parametric plot of the optimized rectification and transport efficiency to the left and (bottom right) the corresponding product as a function of the resonators coupling with $\Delta_{RL}=0$.}\label{Figure3}
\end{figure}

The analysis is further completed by addressing the transport efficiency and its relation to the rectification factor. 
In Fig.~(\ref{Figure3}) we show the figures of merit given by the products $\mathcal{R}T_{L(R)}$, which characterize 
the total diode efficiency while optimizing over the input frequency.
As expected, we can see that the rectification is small for small nonlinearity in the left site. For nonlinearities comparable 
to the dissipation rate, the system presents a left rectification factor of about 0.3, and the corresponding transport efficiency 
increases with the tunnel coupling, $J$, until the system becomes generically a good conductor and the rectification factor 
tends to decrease. Therefore, there is a minimum  of the product $\mathcal{R}T_{L}$, corresponding to the highest rectification 
at highest transport efficiency. Increasing the nonlinearity leads to products of about 0.5, which is ultimately limited by the 
rectification factor. As shown in Fig.~(\ref{Figure3}), this device may achieve almost perfect transport together with a 0.5 
rectification factor.
As expected from the previous analysis, the best diode efficiency is reached when the input frequency is resonant with the 
nonlinear (left) resonator.
Analogously, we show in Fig.~(\ref{Figure3}) the optimization of the right diode efficiency, $\mathcal{R}T_R$. In this case 
the rectification approaches unity with increasing $U/\gamma$, however the corresponding transport is typically smaller 
than 0.2. This amounts to lower, however significant, diode efficiency of the order of 0.15. in this case the maximum efficiency 
is obtained when the laser frequency approximately matches the detuning induced by the nonlinearity $\Delta_L\approx U/2$.
It is interesting that there is a trade-off between transport efficiency and rectification, and in a sense this is the price paid by the 
versatility of the junction.

\begin{figure}	
\includegraphics[width=\linewidth]{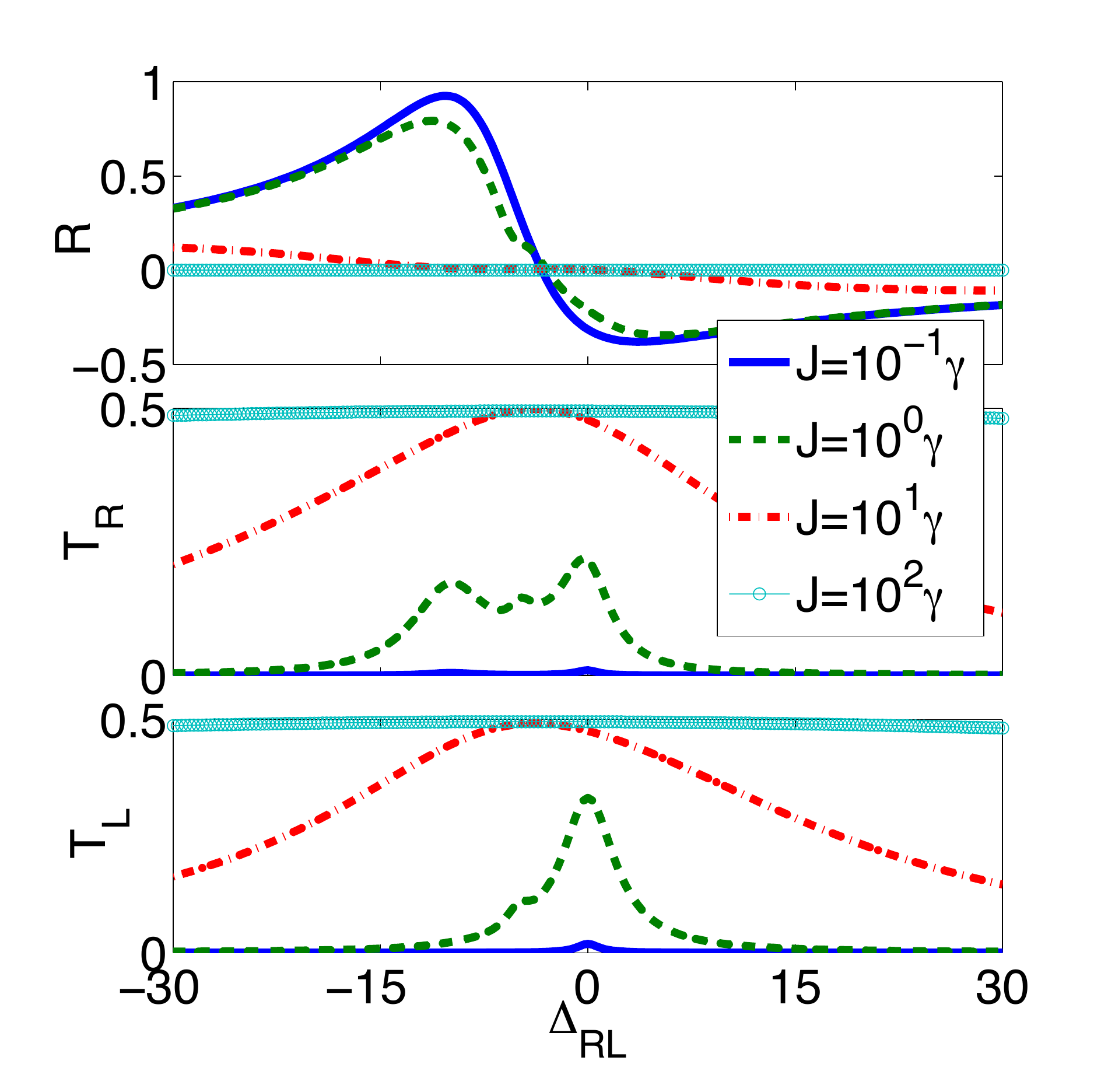}
\caption{{\bf Transport of Fock state $|2\rangle$.} A fast time-dependent pulse can be designed to prepare the Fock state, then the system is allowed to evolve and relax while the output currents are time-integrated, defining the rectification factor (top panel), the transport efficiency to the right (middle), and the transport efficiency to the left (bottom), with $U=10\gamma$ and for different values of the tunnel coupling between resonators. As the coupling is increased the junction slowly switches from a rectifier to a splitter.
}\label{Figure4}
\end{figure}

Now we address the behavior of the junction when it receives a quantum state as an input, instead of a coherent field. 
In practice, this can be achieved by designing an incoming pulse that prepares a pure Fock state in one of the resonators 
with high fidelity. 
Since the Kerr nonlinearity is only activated by two quanta (or higher) Fock states, the n-l junction is reciprocal for a single 
photon Fock state as an input. 
Thus, we study the case of the $|n=2\rangle$ state, such that one photon activates the nonlinearity while the other can be 
rectified. In Fig.~(\ref{Figure4}) we show the rectification and transport efficiency as functions of the resonators detuning, 
for different values of the resonators tunnel coupling. 
Similarly to the continuous wave pumping scenario, we find regimes of left and right rectification. 
Maximum left rectification is found when the resonators are very close to resonance, while maximum right rectification is found 
when the detuning is compensated by the nonlinearity $U-\Delta_{RL}\approx0$. 
Once again we observe a trade-off between rectification and transport efficiency as we increase the resonators coupling, with a 
maximum diode efficiency in the interval $10^0<J<10^1$. 
This trade-off yields an interesting effect in regimes of strong coupling. 
In fact, at large $J$ the junction splits the initial 2-photon Fock state into two distinct wave packets that travel in opposite directions,
irrespective of the direction of the incoming pump pulse, which is indicated by the 0.5 transport efficiency in both directions.

The main challenges to any feasible implementation of the present proposal rely on the system parameters that can be realistically 
achieved in order to observe quantum diode operation, and the detectability of the second-order correlation signals. 
For the first, we refer to the ratio $U/ \gamma$ as the relevant figure of merit, where $\gamma$ is directly related to the resonators 
quality factors through the obvious relation $Q= \omega / \gamma$. 
We point out two different architectures that could be used for the implementation, where highly nonlinear and high quality 
resonators can be fabricated: Superconducting microwave circuits with microstrip (or coplanar) transmission line cavities coupled to 
superconducting qubits \cite{houck_review}, also referred to as \textit{circuit quantum electrodynamics}, 
and semiconducting optical circuits, such as photonic crystal circuits in thin semiconductor slabs \cite{phc_slab_book}.

\textbf{Superconducting microwave circuits.} Recently, tremendous progress has been made in the field of microwave photonic circuits employing superconducting almost-dissipationless elements, such as microstrip transmission line cavities and superconducting qubits \cite{houck_review}, also referred to as \textit{circuit quantum electrodynamics}. In the regime of strong light-matter coupling between a single qubit and a single mode of the superconducting resonator, the system nonlinearity is effectively described by a single-mode Bose-Hubbard model with an effective nonlinearity $U \sim 1$ MHz \cite{hoffmann}. With state-of-the art capabilities, the regime of quantum optical diode operation can be achieved in standard coplanar superconducting resonators with quality-factors on the order of $Q\simeq 10^{5}$, i.e. $\gamma \sim 100$ kHz at microwave frequencies (10 GHz) \cite{hoffmann}, which is enough to reach the condition $U/ \gamma \sim 10$. Moreover, the superconducting microstrip platform naturally realizes the excitation scheme that we have been schematically considering: input/output channels can be defined as broad-band transmission lines of microwave photons directly pumping the n-l junction at left/right ends. Finally, detection of the second-order correlation signal at zero-time delay is now possible also in the  microwave domain through quadrature detection schemes \cite{Lang2011}, which makes it possible for an experimental replication of our theoretical results. 

\textbf{Semiconducting optical circuits.} On a parallel route, a quantum optical diode operation can be realized in integrated photonic circuits at optical or near-infrared wavelengths ($\lambda \sim 1$ $\mu$m). In this case, a preferred platform would be represented by photonic crystal circuits in thin semiconductor slabs \cite{phc_slab_book}. Strong optical nonlinearities of the Kerr-type, which would be the route to effectively realizing the model, have been shown for polaritonic excitations in pillar micro cavities to be on the order of $U_{nl} \sim 9$ $\mu$eV$\cdot \mu$m$^2$~\cite{Ferrier2011}. Diffraction-limited light confinement provided by photonic crystal cavities, i.e. an effective mode area of $(\lambda/n)^2$, would allow to achieve single-photon nonlinearities in the range $U\sim 10 - 100$ $\mu$eV. Considering optical/near-infrared operation, i.e. in the eV range, a quality factor on the order of $10^5 - 10^6$ would be sufficient to reach the quantum optical diode operation $U/ \gamma \sim 10$. We notice that such values have been already demonstrated in typical semiconductor photonic crystal chips \cite{derossi08apl}, although no conclusive signature of single-photon nonlinear behavior has been reported to date for polaritons confined in photonic crystal cavities. On the other hand, such photonic crystal platforms naturally allow to engineer waveguide-coupled cavity schemes, which are straightforwardly described by the theoretical modeling analyzed in the present work. 
It should also be noted that efficient measurement of second-order correlation signals at zero-time delay is achieved through fast single-photon counting at near-infrared wavelengths, where the main limiting factor might be related to the photon lifetime in the resonators, $\tau \sim 1/ \gamma$. For $Q >10^5 $ such lifetime is certainly above the typical resolution timescale of photodetectors (in the pico-second range \cite{hommel2009}), thereby allowing to identify the single- or two-photon rectification regimes in HBT measurements. 

  \textit{Summary---}
We have addressed non-reciprocal propagation of energy pumped into a generic system of tunnel-coupled nonlinear-linear resonators. The model considered has been shown to describe several physical systems. In particular, we have carefully verified that our results could be observed in state-of-the art experiments, and we pointed out two main architectures in which this goal could be pursued.
From a theoretical point of view, we have addressed the quantum nonlinear regime of the proposed device, and we have shown it can work as a rectified quantum source.
The ultimate goal in this research field would be to achieve complete control over quantum optical transport, which includes perfect quantum state transfer and rectification. Taking initial steps in this direction, we have also shown how quantum states at the input may be rectified through the junction, which opens up the possibility to work towards fully 
quantum state rectification, a goal that has never been achieved so far.
Finally, we have shown that the junction also works as a photonic splitter, which shows the versatility of this elementary system as a valve for quantum optical transport. In this respect, a fascinating venue for future research would be the rectification of many-body or mesoscopic quantum states, which would allow for the controlled transport of large amounts of quantum data encoded in complex quantum systems.
In fact, we believe this novel quantum device might become a key element in prospective quantum photonic circuits, where unwanted feedback caused by 
reflections between different system components might have a deleterious effects on the quantum operations to be performed in complex optical networks.

\acknowledgements
The authors thank useful discussions with Emmanuel Pereira. 
E.M. and M.F.S. acknowledge financial support from CNPq, Fapemig, CAPES and INCT-IQ from CNPq. D.V. acknowledges financial support from CNPq.
D.G.acknowledges the Italian Ministry of University and Research through Fondo Investimenti Ricerca di Base 
(FIRB) ÒFuturo in RicercaÓ, Project No. RBFR12RPD1. 
A.A. acknowledges the Fondation Nanosciences de Grenoble.


\begin{thebibliography}{100}


\bibitem{haus_book}
H. A. Haus, \textit{Waves and Fields in Optoelectronics}
(Prentice-Hall, Englewood Cliffs, NJ, 1984).


\bibitem{firstTD}
M. Terraneo, M. Peyrard, and G. Casati,
{Phys. Rev. Lett.} {\bf 88}, 094302 (2002).


\bibitem{chang2006sci}
C.W. Chang, D. Okawa, A. Majumdar, and A. Zettl,
{Science} {\bf 314}, 1121 (2006).

\bibitem{scheibner}
R. Scheibner, M. K\"{o}nig, D. Reuter, A. D. Wieck, C. Gould, H. Buhmann, and L. W. Molenkamp,
{New J. Phys.} {\bf 10}, 083016 (2008).



\bibitem{BLiang2009prl}
B. Liang, B. Yuan, and J. C. Cheng,
{Phys. Rev. Lett.} 103, 104301 (2009).

\bibitem{BLiang2010nmat}
B. Liang, X.-S. Guo, J. Tu, D. Zhang, and J. C. Cheng,
{Nat. Materials} {\bf 9}, 989 (2010).

\bibitem{gallo2001}
K. Gallo, G. Assanto, K. Parameswaran, and M. Fejer,
{Appl. Phys. Lett.} {\bf 79}, 314 (2001).

\bibitem{yu2009nphot}
Z. Yu and S. Fan,
{Nat. Photonics} {\bf 3}, 91 (2009).

\bibitem{Bi2011nphot} 
L. Bi, J. Hu, P. Jiang, D. H. Kim, G. F. Dionne, L. C. Kimerling, and C. A. Ross,
{Nat. Photonics} {\bf 5}, 758 (2011).

\bibitem{Lifan2012} 
L. Fan, J. Wang, L.T. Varghese, H. Shen, B. Niu, Y. Xuan,  A.M. Weiner, and M. Qi,
{Science} \textbf{335}, 447 (2012).

\bibitem{Lira2012} 
H. Lira, Z. Yu, S. Fan, and M. Lipson,
{Phys. Rev. Lett.} \textbf{109}, 033901 (2012).

\bibitem{qp_review}
J.L. O'Brien, A. Furusawa, and J. Vu\v{c}kovi\'{c},
{Nat. Photonics} {\bf 3}, 687 (2009).

\bibitem{chang07np}
D.E. Chang, A.S. Sorensen, E.A. Demler, and M.D. Lukin,
Nat. Physics {\bf 3}, 807 (2007).

\bibitem{gerace_josephson}
D. Gerace, H.E. T\"{u}reci, A. Imamo\v{g}lu, V. Giovannetti, and R. Fazio,
{Nat. Physics} {\bf 5}, 281 (2009).


\bibitem{shen2011prl} 
Y. Shen, M. Bradford, and J.-T. Shen,
{Phys. Rev. Lett.} \textbf{107}, 173902 (2011).

\bibitem{AtDiode} 
R. A. Pepino, J. Cooper, D. Z. Anderson, and M. J. Holland, 
Phys. Rev. Lett. {\bf 103}, 140405 (2009). 



\bibitem{casatiGPD} 
S. Lepri and G. Casati,
{Phys. Rev. Lett.} {\bf 106}, 164101 (2011).











\bibitem{BHtheory} 
D. Jaksch, C. Bruder, I. J. Cirac, C. W. Gardiner, and P. Zoller,
{Phys. Rev. Lett.} {\bf 81}, 3108 (1998).

\bibitem{BHexperiment} 
M. Greiner, O. M. Mandel, T. Esslinger, T. Hansch, and I. Bloch,
{Nature}  {\bf 415}, 39 (2002). 





\bibitem{Imamoglu}  
M.J. Werner and A. Imamoglu. 
{Phys. Rev. A} \textbf{61}, 011801(R) (1999).

\bibitem{hartmann06}
M. J. Hartmann, F. G. S. L. Brand\~ao, and M. B. Plenio,
{Nat. Physics} {\bf 2}, 849 (2006).

\bibitem{chang08np}
D.~E. Chang, V. Gritsev, G. Morigi, V. Vuletic, M.~D. Lukin, and E.~A. Demler,
{Nat. Physics} {\bf 4}, 884 (2008).

\bibitem{angelakis07}
D.~G. Angelakis, M.~F. Santos, and S.~Bose,
{Phys. Rev. A} {\bf 76}, 031805(R) (2007).

\bibitem{leib2010njp}
M. Leib and M. J. Hartmann,
{New J. Phys.} {\bf 12}, 093031 (2010).

\bibitem{carusotto2009}
I. Carusotto, D. Gerace, H. E. Tureci, S. De Liberato, C. Ciuti, and A. Imamoglu,
{Phys. Rev. Lett.} {\bf 103}, 033601 (2009).

\bibitem{nissen2012}
F. Nissen, S. Schmidt, M. Biondi, G. Blatter, H. E. Tureci, and J. Keeling,
{Phys. Rev. Lett.} \textbf{108}, 233603 (2012).

\bibitem{ciuti06prb}
A. Verger, C. Ciuti, and I. Carusotto,
{Phys. Rev. B} {\bf 73}, 193306 (2006).

\bibitem{carusot2010epl}
I. Carusotto, T. Volz, and A. Imamo\v{g}lu.
{Europhys. Lett.} {\bf 90}, 37001 (2010).   

\bibitem{ferretti2012} 
S. Ferretti and D. Gerace,
{Phys. Rev. B} {\bf 85}, 033303 (2012).

\bibitem{savona10prl}
T.C.H. Liew and V. Savona, 
Phys. Rev. Lett. {\bf 104}, 183601 (2010).

\bibitem{bamba}
M. Bamba, A. Imamo\v{g}lu, I. Carusotto, and C. Ciuti,
Phys. Rev. A {\bf 83}, 021802(R) (2011).

\bibitem{ferretti2013}
S. Ferretti, V. Savona, and D. Gerace,
New J. Phys. {\bf 15}, 025012 (2013).






\bibitem{QMaC} 
H. Wiseman and G. Milburn, \textit{Quantum Measurement and Control} (Cambridge, 2010).

\bibitem{QNoise} C. Gardner, \textit{Quantum Noise} (Springer, 2010). 

\bibitem{carmichael_book}
H. Carmichael,  \textit{An open systems approach to quantum optics}
(Springer-Verlag, Berlin, 1993).

\bibitem{hbt_original}
R. Hanbury Brown \& R.Q. Twiss.
\textit{Nature}  {\bf 177}, 27 (1956). 


\bibitem{houck_review}
A.A. Houck, H.E. Tureci, and J. Koch,
{Nat. Physics} {\bf 8}, 292 (2012).

%
%

\bibitem{phc_slab_book}
J. D. Joannopoulos, S. G. Johnson, J. N. Winn, and R. D. Meade, 
\textit{Photonic Crystals: molding the flow of light} (Princeton University Press, Princeton, 2008).


\bibitem{hoffmann}
A.J. Hoffman, S.J. Srinivasan, S. Schmidt, L. Spietz, J. Aumentado, H.E. Tureci, \& A.A. Houck.
\textit{Phys. Rev. Lett.} {\bf 107}, 053602 (2011).

\bibitem{Lang2011}
C. Lang, D. Bozyigit, C. Eichler, L. Steffen, J.M. Fink, A.A. Abdumalikov, M. Baur, S. Filipp, M.P. da Silva, A. Blais, \& A. Wallraff.
\textit{Phys. Rev. Lett.} {\bf 106}, 243601 (2011).

\bibitem{Ferrier2011}
L. Ferrier, E. Wertz, R. Johne, D. D. Solnyshkov, P. Senellart, I. Sagnes, A. Lamaitre, G. Malpuech, \& J. Bloch.
\textit{Phys. Rev. Lett.} \textbf{106}, 126401 (2011).


\bibitem{derossi08apl}
Combri\'e, S., De Rossi, A., Tran, Q. V. \& Benisty, H.
{\it Opt. Letters} $\mathbf{33}$, 1908-1910 (2008).

\bibitem{hommel2009}
J. Wiersig, C. Gies, F. Jahnke, M. Assmann, T. Berstermann, M. Bayer, C. Kistner, S. Reitzenstein, C. Schneider,
S.Hofling, A. Forchel, C. Kruse, J. Kalden, \& D. Hommel.
\textit{Nature}  {\bf 460}, 245 (2009).

%
%
%

\end{thebibliography}
\end{document}